\documentclass{emulateapj}
\slugcomment{{\sc Submitted to ApJL:} \today} 
\usepackage{natbib}
\usepackage{url}

\begin{document}
\title{The Long-Lived UV ``Plateau" of SN 2012aw}

\author{Amanda J. Bayless\altaffilmark{1}, Tyler Pritchard\altaffilmark{2}, Peter W. A. Roming\altaffilmark{1, 2}, Paul Kuin\altaffilmark{3}, Peter J. Brown\altaffilmark{4}, Maria Teresa Botticella\altaffilmark{5}, Massimo Dall'Ora\altaffilmark{5}, Lucille H. Frey\altaffilmark{6, 7}, Wesley Even\altaffilmark{6}, Chris L. Fryer\altaffilmark{6, 8, 9}, Justyn R. Maund\altaffilmark{10, 11, 12}, Morgan Fraser\altaffilmark{10}}

\altaffiltext{1}{Southwest Research Institute, Department of Space Science, 6220 Culebra Rd, San Antonio, TX 78238, USA}
\altaffiltext{2}{Department of Astronomy \& Astrophysics, Penn State University, 525 Davey Lab, University Park, PA 16802, USA}
\altaffiltext{3}{Mullard Space Science Laboratory, Holmbury St. Mary, Dorking, Surrey, RH5 6NT, UK}
\altaffiltext{4}{Texas A\&M University, Department of Physics and Astronomy, College Station, TX 77843-4242, USA}
\altaffiltext{5}{INAF-Osservatorio Astronomico di Capodimonte, via Moiariello 16, 80131, Napoli, Italy.}
\altaffiltext{6}{Los Alamos National Laboratory, Los Alamos, NM 87545, USA}
\altaffiltext{7}{Department of Computer Science, University of New Mexico, Albuquerque, NM  87131, USA}
\altaffiltext{8}{Physics Department, University of Arizona, Tucson, AZ 85721, USA}
\altaffiltext{9}{Physics and Astronomy Department, University of New Mexico, Albuquerque, NM 87131, USA}
\altaffiltext{10}{Astrophysics Research Center, School of Mathematics and Physics, Queen's University Belfast, Belfast BT7 1NN, UK}
\altaffiltext{11}{Dark Cosmology Centre, Niels Bohr Institute, University of Copenhagen, Juliane Maries Vej, DK-2100 Copenhagen, Denmark}
\altaffiltext{12}{Royal Society Research Fellow}

\begin{abstract}
Observations with the {\em Swift} UVOT have unambiguously uncovered for the first time a long-lived, UV ``plateau" in a Type II-P supernova (SN).
  Although this flattening in slope is hinted at in a few other SNe, due to its proximity and minimal line-of-sight extinction, SN 2012aw has afforded the first opportunity to clearly observe this UV plateau.  The observations of SN 2012aw revealed all {\em Swift} UV and {\it u}-band lightcurves initially declined rapidly, but 27 days after explosion the light curves flattened.  Some possible sources of the UV plateau are: the same thermal process that cause the optical plateau, heating from radioactive decay, or a combination of both processes.
\end{abstract}
\keywords{supernovae: general---supernovae: individual (SN 2012aw)}
\section{Introduction}
A Type II-P supernova (SN) is the most common type of SNe in a volume limited sample (\citealt{li11} and references therein) and results from the core collapse of a massive star, usually a red super giant (RSG).  The gravitational binding energy is approximately $10^{53}$ ergs, but only a small percentage of this is eventually released as sustained emission with a life time of a few months.  The luminosity in the optical is maintained by the recombination of hydrogen and excitation by radioactive isotopes even as the expanding photosphere cools.  The initial emission is very blue, with the peak of the bolometric luminosity in the UV, but as the photosphere expands and cools, it becomes redder \citep{kir73,kir75,mit02,leo02,bro07}.  Thus, the optical light curve often remains at nearly constant brightness (plateaus) for several weeks.  

In spite of the copious number of Type~II-P SNe, very few have been well studied in the UV since this requires space-based instruments and a nearby event with low extinction.  Recent UV observations have been carried out by the {\em Swift} \citep{geh04} UV Optical Telescope (UVOT; \citealt{rom00,rom04,rom05}), IUE (SN 1987A; \citealt{pun95}), HST (SN 1999em; \citealt{bar00}), and GALEX (SN 2005ay; \citealt{gal08}).    These SNe are UV bright early on and can be seen at great distances, but the UV flux usually fades after a few weeks as the photosphere cools \citep{bro09}.  For most distant SNe this means that the late-time UV flux is too faint to be observable. In the rare cases where the SNe were visible in the UV at late epochs, there are indications of a change to a flatter slope, but there is little data at these late times to determine anything definitive.\footnote[1]{SN 1987A was extensively observed, but it was also peculiar.}  This has changed with the advent of SN 2012aw.

SN 2012aw was discovered on 2012 March 16.9 \citep{fag12} in the nearby galaxy M95 (10 Mpc; \citealt{fre01}).  M95 was observed on March 15.3 \citep{poz12} without a detection to a limiting magnitude of $R > 20.7$.  From this constraint, \citet{fra12} adopt an explosion date of 2012 March 16.0 $\pm$ 0.8d (JD 2456002.5), which we adopt as Day 0 throughout this paper.  Pre-explosion images of M95 in the region of SN 2012aw indicate an IR bright star in the vicinity, likely a RSG \citep{fra12,van12}.  Follow up spectra \citep{mun12, ito12, siv12} identified it as a core-collapse, with a preliminary classification as a Type~II-P. The plateau classification is confirmed by the optical photometry from {\em Swift} as shown in this paper. The nearness and minimal total extinction (line-of-sight: $A_v = 0.37$, host: $A_v = 0.24$, with $R_v=3.1$; \citealt{van12}) of SN 2012aw meant that unprecedented UV observations of a SN covering the first 111 days past explosion, could be obtained.  We also observed this SN with the UVOT  UV grism for approximately two weeks post-explosion, before it faded beyond a reasonable signal-to-noise threshold.  

As in prior Type~II-P SNe, SN 2012aw is initially UV bright followed by a rapid decline in flux.  At about Day 27 the UV light curve slope significantly flattened to a very shallow sloped decline, a UV ``plateau." This UV plateau continued though to Day 111, at which time {\em Swift} could no longer observe the SN due to observing constraints.   In this paper we present the {\em Swift} optical and UV light curves.  We also present and compare the time-series UV spectra of SN 2012aw to previously observed spectral models of Type II-P SNe (SN 2005cs and SN 2006bp) and present possible causes of the UV flattening.

\section{Observations and Data Reduction}
\subsection{{\it Swift} UVOT Photometry}

 \begin{table*}[htb]
 {\scriptsize
 \center
 	 \caption{Observations of SN 2012aw from UVOT.}
 	\label{tab1}
   \begin{tabular}{cccccccc}
\hline
&&Photometric & Magnitudes$^a$ \\
JD (+2450000) & {\it uvw2}  & {\it uvm2} 		& {\it uvw1} 		& 	{\it u} 		& 	{\it b} 		& {\it v} \\
  \hline
6006.2	&	12.20$\pm$0.22	&	12.09$\pm$0.32	&	12.16$\pm$0.12	&	12.32$\pm$0.08	&	13.56$\pm$0.08	&	13.71$\pm$0.07	\\
6008.0	&	12.83$\pm$0.29	&	...				&	...				&	...				&	...				&	...	\\
6009.8	&	...				&	...				&	12.64$\pm$0.09	&	...				&	...				&	...	\\
6009.9	&	13.34$\pm$0.22	&	...				&	...				&	...				&	...				&	...	\\
6011.9	&	...				&	...				&	{\it 12.56}		&	...				&	...				&	...	\\
6012.2	&	13.77$\pm$0.22	&	...				&	...				&	...				&	...				&	...	\\
6012.5	&	...				&	...				&	...				&	12.47$\pm$0.14	&	...				&	...	\\
6012.8	&	...				&	...				&	12.90$\pm$0.12	&	...				&	...				&	...	\\
6014.0	&	14.16$\pm$0.25	&	14.01$\pm$0.21	&	13.17$\pm$0.21	&	12.52$\pm$0.12	&	13.53$\pm$0.11	&	13.50$\pm$0.10	\\
6015.2	&	...				&	14.41$\pm$0.20	&	...				&	...				&	...				&	...	\\
6015.5	&	14.52$\pm$0.11	&	...				&	...				&	...				&	...				&	...	\\
6015.9	&	14.63$\pm$0.21	&	14.64$\pm$0.09	&	13.56$\pm$0.19	&	12.60$\pm$0.12	&	13.55$\pm$0.11	&	13.53$\pm$0.10	\\
6017.0	&	14.91$\pm$0.17	&	...				&	...				&	12.71$\pm$0.27	&	...				&	...	\\
6018.0	&	15.18$\pm$0.16	&	15.30$\pm$0.12	&	14.04$\pm$0.14	&	12.80$\pm$0.12	&	13.58$\pm$0.10	&	13.50$\pm$0.09	\\
6018.6	&	15.39$\pm$0.10	&	15.68$\pm$0.08	&	...				&	...				&	...				&	...	\\
6019.9	&	15.87$\pm$0.14	&	16.10$\pm$0.10	&	14.55$\pm$0.14	&	13.04$\pm$0.14	&	13.65$\pm$0.11	&	13.46$\pm$0.11	\\
6024.0	&	17.00$\pm$0.11	&	17.64$\pm$0.09	&	15.53$\pm$0.11	&	13.76$\pm$0.14	&	13.79$\pm$0.13	&	13.45$\pm$0.11	\\
6025.9	&	17.29$\pm$0.11	&	18.17$\pm$0.10	&	15.93$\pm$0.11	&	14.08$\pm$0.14	&	13.86$\pm$0.14	&	13.45$\pm$0.13	\\
6028.1	&	17.50$\pm$0.11	&	18.36$\pm$0.10	&	16.10$\pm$0.09	&	14.39$\pm$0.12	&	13.96$\pm$0.12	&	13.46$\pm$0.11	\\
6029.0	&	17.56$\pm$0.10	&	...				&	...				&	14.56$\pm$0.31	&	...				&	...	\\
6029.6	&	17.61$\pm$0.10	&	...				&	...				&	...				&	...				&	...	\\
6029.7	&	17.72$\pm$0.12	&	...				&	...				&	...				&	...				&	...	\\
6030.7	&	17.72$\pm$0.10	&	18.94$\pm$0.13	&	16.53$\pm$0.08	&	14.77$\pm$0.10	&	14.09$\pm$0.11	&	13.51$\pm$0.10	\\
6031.9	&	17.78$\pm$0.10	&	...				&	16.53$\pm$0.09	&	...				&	...				&	...	\\
6032.7	&	17.86$\pm$0.13	&	...				&	...				&	14.82$\pm$0.09	&	...				&	...	\\
6033.8	&	17.98$\pm$0.11	&	19.33$\pm$0.16	&	16.65$\pm$0.08	&	14.97$\pm$0.09	&	14.19$\pm$0.10	&	13.54$\pm$0.09	\\
6035.8	&	18.11$\pm$0.11	&	19.44$\pm$0.15	&	16.87$\pm$0.09	&	15.20$\pm$0.11	&	14.27$\pm$0.13	&	13.60$\pm$0.12	\\
6039.6	&	...				&	19.52$\pm$0.16	&	...				&	...				&	...				&	13.61$\pm$0.10	\\
6039.7	&	18.29$\pm$0.11	&	...				&	16.95$\pm$0.09	&	15.45$\pm$0.10	&	14.36$\pm$0.13	&	...	\\
6043.7	&	18.48$\pm$0.12	&	19.62$\pm$0.19	&	17.09$\pm$0.09	&	15.72$\pm$0.08	&	14.46$\pm$0.10	&	13.62$\pm$0.09	\\
6045.7	&	18.42$\pm$0.11	&	19.66$\pm$0.17	&	17.19$\pm$0.09	&	15.76$\pm$0.09	&	14.52$\pm$0.12	&	13.64$\pm$0.11	\\
6046.7	&	18.50$\pm$0.11	&	19.91$\pm$0.20	&	17.28$\pm$0.09	&	15.90$\pm$0.09	&	14.54$\pm$0.13	&	13.64$\pm$0.11	\\
6050.9	&	18.63$\pm$0.12	&	20.19$\pm$0.23	&	17.37$\pm$0.09	&	16.14$\pm$0.09	&	14.64$\pm$0.13	&	13.66$\pm$0.12	\\
6052.0	&	18.60$\pm$0.12	&	19.83$\pm$0.22	&	17.38$\pm$0.09	&	16.12$\pm$0.08	&	14.69$\pm$0.10	&	13.70$\pm$0.09	\\
6054.2	&	18.47$\pm$0.11	&	20.08$\pm$0.22	&	17.51$\pm$0.09	&	16.20$\pm$0.09	&	14.67$\pm$0.12	&	13.64$\pm$0.11	\\
6057.8	&	18.65$\pm$0.12	&	20.40$\pm$0.28	&	17.61$\pm$0.09	&	16.40$\pm$0.08	&	14.80$\pm$0.11	&	13.68$\pm$0.11	\\
6061.8	&	18.82$\pm$0.12	&	20.53$\pm$0.32	&	17.68$\pm$0.09	&	16.63$\pm$0.08	&	14.84$\pm$0.11	&	13.73$\pm$0.10	\\
6066.3	&	...				&	...				&	17.88$\pm$0.16	&	...				&	...				&	...	\\
6074.1	&	19.30$\pm$0.16	&	20.26$\pm$0.29	&	17.82$\pm$0.10	&	17.02$\pm$0.09	&	15.03$\pm$0.09	&	13.79$\pm$0.09	\\
6078.2	&	18.92$\pm$0.13	&	20.33$\pm$0.29	&	17.92$\pm$0.10	&	17.07$\pm$0.09	&	15.11$\pm$0.10	&	13.83$\pm$0.09	\\
6113.4	&	19.67$\pm$0.20	&	{\it 20.70}		&	18.42$\pm$0.13	&	18.11$\pm$0.15	&	15.66$\pm$0.08	&	14.22$\pm$0.08	\\
\hline
\hline
UV GRISM &Start Time  		&  Stop Time  	 & Start Time    		&  Stop Time  	 & exposure 	\\
Number & (UT) &  (UT) &  (JD) 	&  (JD) & 		(sec)			\\
\hline
1&2012-03-21T00:16:24 & 2012-03-21T23:03:24 & 2456007.51          & 2456008.46          &  8287.4   \\
2&2012-03-23T00:41:04 & 2012-03-24T21:42:08 & 2456009.52          & 2456011.40          &   5116.7   \\
3&2012-03-25T00:48:24 & 2012-03-26T01:09:46 & 2456011.53          & 2456012.55          &   8021.7  \\
4&2012-03-28T08:47:25 & 2012-03-28T23:43:28 & 2456014.87          & 2456015.49          &  5663.9   \\
5&2012-03-30T05:44:24 & 2012-03-30T18:43:24 & 2456016.74          & 2456017.28          &  8978.1  \\
6&2012-04-11T01:42:00 & 2012-04-11T19:45:40 & 2456028.57          & 2456032.72          &  22759.8   \\
\hline
\hline
&&Spectrophotometric & Magnitudes \\
Number & {\it uvw2}$^b$  & {\it uvm2} 		& {\it uvw1} 		& 	{\it u} 		& 	{\it b} 		& {\it v}$^b$ \\
1 &12.979 & 12.494 & 12.348 & 12.327 & 13.335 & 13.019 \\
2 &13.467 & 12.963 & 12.545 & 12.344 & 13.353 & 13.028 \\
3 &14.701 & 13.522 & 12.897 & 12.390 & 13.330 & 12.989 \\
4 &14.867 & 14.520 & 13.566 & 12.560 & 13.369 & 13.063 \\
5 &15.050 & 14.929 & 13.934 & 12.685 & 13.435 & 13.120 \\
6 & -- $^c$ & -- $^c$ & -- $^c$ & 14.649 & 14.059 & 13.482\\
\hline
\end{tabular}
\begin{flushleft}
\footnotesize{(a) Photometry values in italics are 3$\sigma$ upper limits.  All photometric errors are 3$\sigma$.  \\
(b) The spectrophotometric magnitudes for {\it uvw2} and {\it v} are only limits as the UV grism does not fill these bandpasses.\\
(c) The last spectrum blueward of 2940 \AA\ is in the noise level.\\}
\end{flushleft}
}
\end{table*}

SN 2012aw was observed from 2012 March 19 (UT; $\sim$3 days post-explosion) to May 30 (JD 2456006.2 to 2456078.2) with the {\em Swift} UVOT with a cadence of every one to four days. The UVOT observed in six UV/optical filters: {\it uvw2} ($\lambda_c =1928$ \AA), {\it uvm2} ($\lambda_c =2246$ \AA), {\it uvw1} ($\lambda_c =2600$ \AA), {\it u} ($\lambda_c =3465$ \AA), {\it b} ($\lambda_c =4392$ \AA), and {\it v } ($\lambda_c =5468$ \AA; \citealt{poo08}).  Photometry using a $3^{\prime\prime}$ source aperture, including galaxy flux subtraction using a previous {\em Swift} UVOT observation of M95 as a template, was performed following the method outlined in \citet{bro09}. The data reduction pipeline used HEASOFT 6.6.3 and {\em Swift} Release 3.3 analysis tools with UVOT zero-points \citep{poo08} and updated calibrations  \citep{bre11}.  

\begin{figure}[htb]
\center \includegraphics[angle=0,scale=.4]{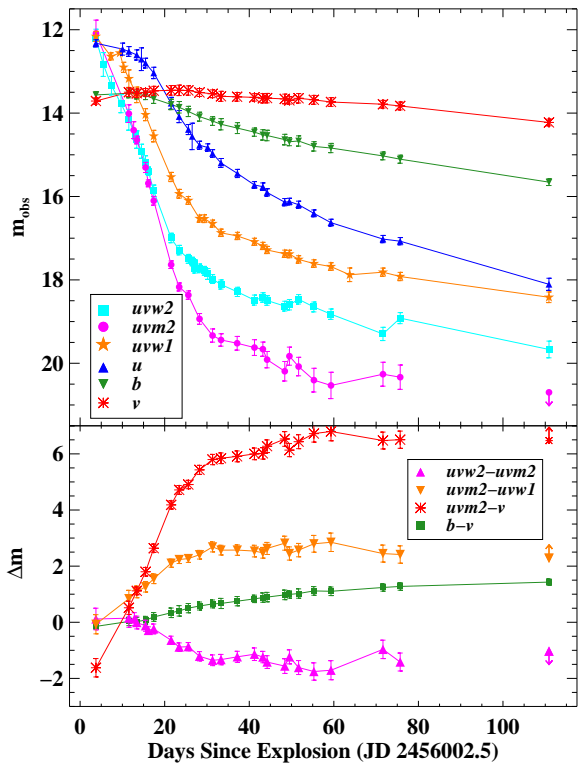}
\caption{{\it Top.} -- The UV and optical light curves of SN 2012aw from {\it Swift} UVOT.  The optical {\it v} and {\it b} bands are typical of a Type~II-P SN.  The UV bands decrease in brightness rapidly, but $\sim 27$ days past explosion the light curve flattens out and follows a similar slope as the optical bands. {\it Bottom.} -- UV and optical colors of SN 2012aw.  The $uvm2-v$ (UV-optical) color curve indicates that the SN becomes quite red over time.}
\label{lc}
\end{figure}

The upper panel of Table \ref{tab1} gives a summary of the photometric observations.  Figure \ref{lc} shows the multi-color light curve and UV/optical colors of SN
2012aw. The {\it v}- and {\it b}-band are largely flat, with declining slopes over the range of observations of $0.006\pm0.001$ and $0.023\pm0.001$ mag/day, respectively ($1\sigma$ errors), the characterizing signature of a Type II-P SN.  The SN is UV bright initially, but declines rapidly with slopes for the {\it uvw2}, {\it uvm2}, and {\it uvw1}, of $0.311\pm0.012$, $0.197\pm0.010$, and $0.257\pm0.007$ mag/day, respectively.  After Day 27, the lightcurves flattened into a slowly declining plateau with slopes very similar to the {\it b}-band of $0.028\pm0.003$, $0.029\pm0.005$, and $0.031\pm0.002$ mag/day, respectively.  The {\it u}-band light curve had two changes in slope.  The initial trend was rather flat, similar to the optical light curves, with a slope of $0.038\pm0.007$ mag/day until about Day 15 (JD 2456018). Then, the {\it u}-band followed the trend of the UV light curves, decreasing in brightness at a rate of $0.166\pm0.004$ mag/day until about Day 27-30, where its light curve showed a slower decline of $0.050\pm0.003$ mag/day.

The {\it uvm2 - v} color overall becomes quite red, as expected as the photosphere cools.  The {\it uvw2 - uvm2} color does become blue initially.  This is because the {\it uvm2} flux declines faster than the {\it uvw2} flux.  Figure \ref{uvspec} shows the time-series UV spectra evolution, which is described in the following section.  The flux near 1900 \AA\ ({\it uvw2}) remains nearly constant as the blackbody peak moves through this waveband and there are not many metal lines at this temperature ($\sim$ 20,000 K; bolometric temperature, see Section 4). Near 2100 \AA\ ({\it uvm2}) in the first spectrum there is a flux peak from the absence of iron lines.  In the later spectra, this is line blanketed most likely by \ion{Ni}{2}.  A similar phenomena happened with SN 1987A \citep{cas87}. 

\subsection{Swift UVOT Spectroscopy}
{\em Swift} UVOT started observing with the UV grism (calibrated for the range $\lambda 1650 - \lambda 4900$ \AA) on 2012 March 21 (UT; $\sim$5 days post-explosion).  Spectra from individual images were summed in order to limit the noise in the spectrum (Figure 2). The bottom panel of Table \ref{tab1} lists the start and end times of each summed observation and the total exposure time. The spectral features are red-shifted due to M95's recessional velocity of 788 km/s.

\begin{figure}
\center \includegraphics[angle=90,scale=.32]{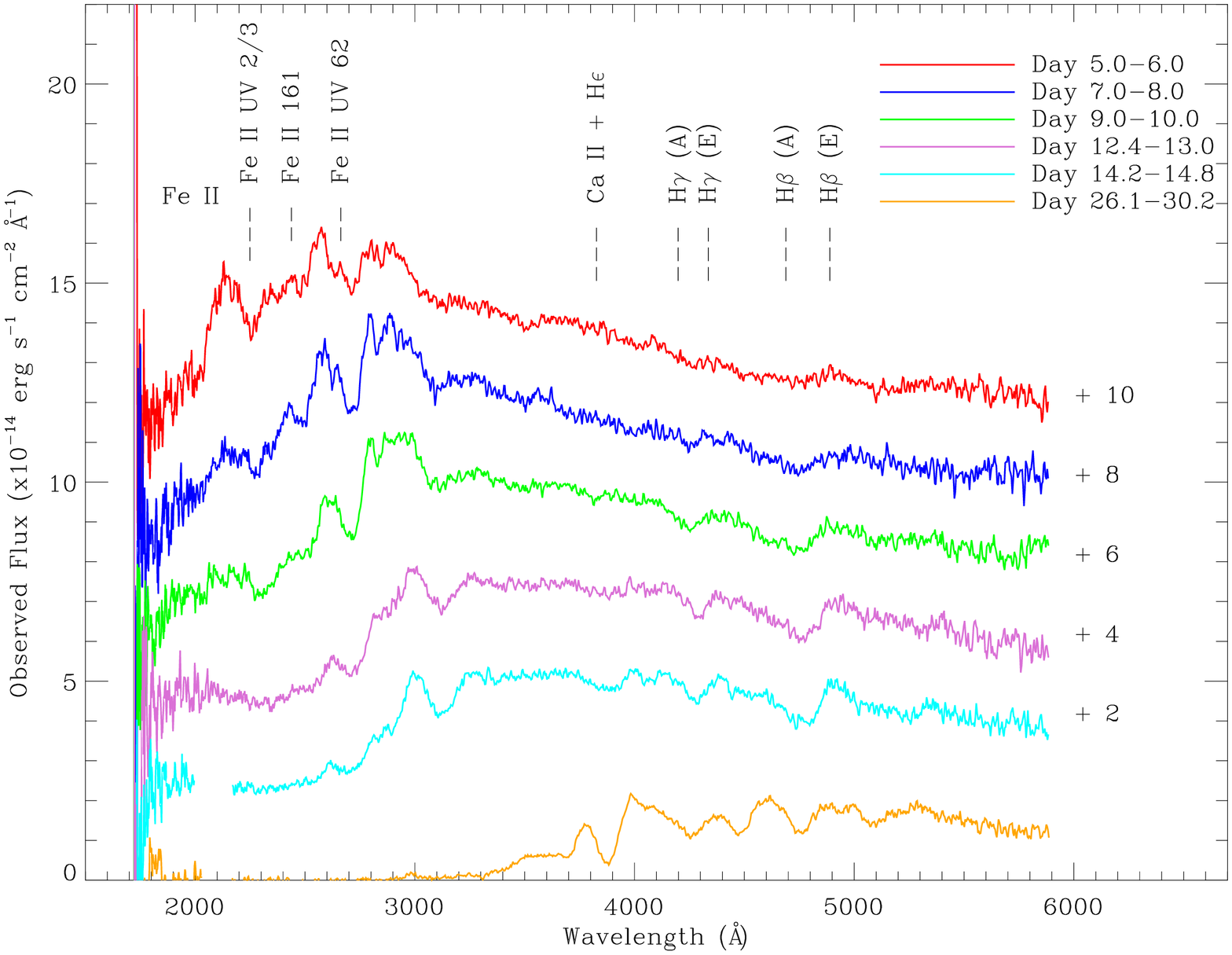}
\caption{Time-Series spectra from the UVOT UV grism.  The spectra are offset from each other by adding a constant amount of flux for clarity.  The individual hydrogen absorption (A) and emission (E) components are also labeled.
\label{uvspec}}
\end{figure}

A feature of the {\em Swift} UV grism is the ability to observe at offsets from the center (default) position.  These positions take advantage of the order 
separation, which is more pronounced when the grism is not at the default position. This means that the blue part of the spectrum can also be observed in second 
order, and the contamination of the first order by the second order is reduced substantially.  All but the last spectrum were observed at offset positions.  The 
offset was moved closer to the central position as time went on. A new preliminary calibration for the offset positions was used (Kuin et al. 2012, in prep.), 
which has an accuracy of 15 \AA\ and is reliable to  within 20\% below 1800 \AA\ and $<$ 10\% for the UV region between 1800 - 4900 \AA.  The spectra were 
extracted following the same procedure as for SN 2011ht \citep{rom12}, using version 0.9.6 of the Python software for UVOT grism analysis \footnote[2]{\url
{http://www.mssl.ucl.ac.uk/{\raise.17ex\hbox{$\scriptstyle\sim$}}npmk/Grism/uvotpy/}}. The last spectrum was observed at the central default position because the 
UV-emission was greatly reduced already, meaning that the blue part of the second order gives minimal contamination to the first order.  The first five spectra 
have second order contamination starting at $\sim$ 4900  \AA\ which is responsible for the upward jump in the flux.   The last two spectra have some 
contamination by a zeroth order between 2030 - 2167 \AA, which was removed, producing the gap in Figure \ref{uvspec}. The background noise level in the spectra 
is $6 \times 10^{-16}$ ergs cm$^{-2}$ s$^{-1}$ \AA $^{-1}$.  The last spectrum blueward of 2940 \AA\ is below this noise level.

To check the calibrations we produced spectrophotometric {\em Swift} magnitudes from the spectra in Figure \ref{uvspec}, which are tabulated in Table \ref{tab1}.  The UV grism does not completely cover the {\it uvw2} and the {\it v}-band and therefore these are only limits. The UV colors from the last spectra are also only limits because of the noise level.  This spectrophotometry is consistent with the photometry to within the errors on the spectra of $\sim 5-10\%$.  Because of this and that the UVOT photometry is will studied, we are confident that the UV plateau is a real feature of the SN 2012aw lightcurve.

\section{Discussion}
\subsection{UV Light Curve}

The UV light curve fell rapidly and then flattened out approximately one month after the explosion.  Unlike many UV filters used, {\it uvm2} has effectively no red leak and therefore we can be confident that the UV plateau is a real feature of the SN. There is a red leak in {\it uvw1} and {\it uvw2}, but applying red leak correction \citep{bro10} still shows the plateau, just $\sim 2.5$ mag fainter in both filters.  The red leak does not contribute at all to the early time lightcurve up to $\sim$ Day 21. 

\begin{figure}[htb]
 \center 
\includegraphics[angle=90,scale=.32]{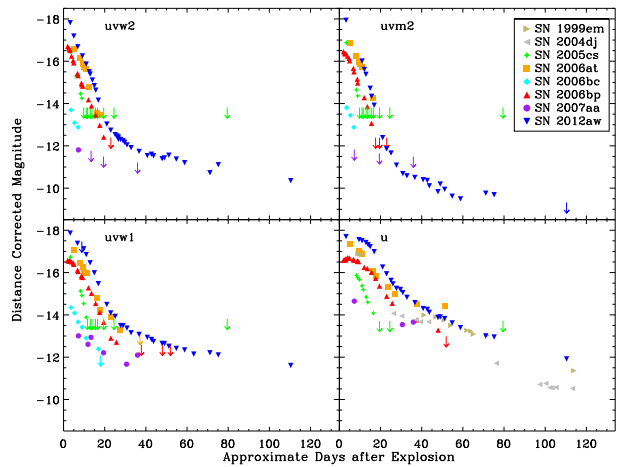}
 \caption{Type~II-P SNe UV and {\it u}-band light curves corrected for distance, but not extinction. The lightcurves from SN 1999em are from \citet{elm03}, SN 2004dj from \citet{tsv08}, SN 2005cs and SN 2006bp from \citet{des08}, and SN 2006at, SN 2006bc, and SN 2007aa from \citet{bro09}. The arrows indicate upper limits. \label{lccomp_multi}} 
\end{figure}

Figure \ref{lccomp_multi} shows the UV and {\it u}-band lightcurve of SN 2012aw along with other Type~II-P SNe corrected for distance, but not extinction since the extinction in the host galaxy for several of these is unknown.  The worst case scenario is likely SN 2006bc, which has a total $E(B-V)=0.52$, corresponding to $\sim4.3$ mag change in {\it uvw2} and {\it uvm2}, $\sim3.5$ mag change in {\it uvw1}, and $\sim2.6$ mag change in {\it u}-band, assuming the extinction of \citet{car89}. This flattening to the UV slope at late times is suggested, but not well observed in the prior UV studies of Type~II-P SNe, as the UV flux fell below detection levels within a few weeks.  We also show the ground-based {\it U}-band for SN 1999em and 2004dj, which do extend past 100 days.  Both of these also show a flat {\it U}-band slope, like SN 2012aw. SN 2012aw's UV light curve appears similar in shape to that of SN 2006at, SN 2006bp and possibly SN 2006bc, although the latter has very few observations.  The UV light curves of SN 2005cs also appear similar if the lightcurves are shifted forward by a few days. 

\subsection{UV Spectral Evolution}
In Figure \ref{uvspec} the {\em Swift} UV spectra are plotted. Since the UVOT is a photon-counting instrument, the flux errors are well understood. They are smallest around 3000 \AA, where the sensitivity is highest, and largest below 1800 \AA\ and above 4000 \AA. The brightness of the SN is low enough that effects of coincidence loss \citetext{\citealp{poo08}; \citealp{kui08}; Kuin et al. 2012, in prep.} are small and dominated by the coincidence loss in the background \citep{bre10}.  These UV spectra are the most detailed UV spectra of an early Type~II-P SNe. The final spectrum taken two weeks after explosion still shows some near-UV emission. 

In Figure \ref{uvspec}, the main spectral features are identified. The main spectral absorption minima are blue-shifted by 6000-8000 km/s, while the edge velocities in H$\beta$ and H$\gamma$ P-Cygni profiles are $\sim$15,000 and $\sim$13,000 km/s.  The UV end of the spectrum decreases in brightness rapidly over the course of the month.  This decrease is likely due to iron line blanketing with the addition of nickel after $\sim$ 10 days

\begin{figure}[htb]
 \center \includegraphics[angle=90,scale=.32]{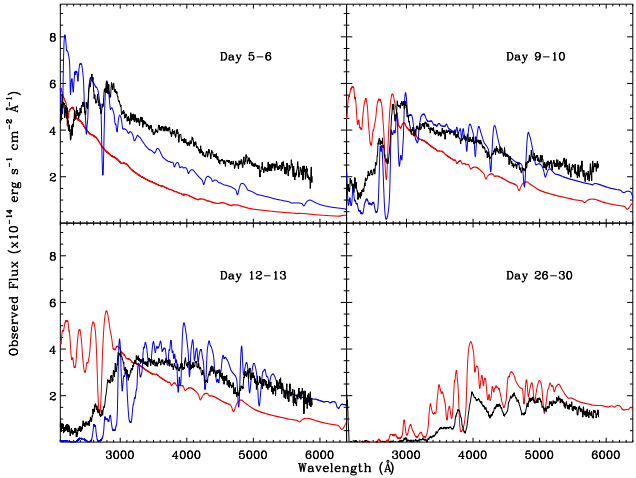}
 \caption{UV time-series spectrum of SN 2012aw (black) with the spectral models \citep{des08} for SN 2005cs (blue) and SN 2006bp (red) scaled by eye to the SN 2012aw observations and extinction applied using the reddening to SN 2012aw.}
 \label{uvmodel}
\end{figure}

Figure \ref{uvmodel} shows the  SN 2012aw observed spectra along with two spectral models for nearly the same post-explosion time.  The models are from \citet{des08} and were originally created from the analysis of {\em Swift} UVOT data of SN 2005cs and SN 2006bp.  The model parameters of SN 2006bp and 2005cs (given in Table 6 and Table 7 of \citealt{des08}) have been extinction corrected from the empirical extinction formula of \citet{car89} using the same Milky Way reddening for SN 2012aw of $E(B-V)=0.025$ mag \citep{sch11}. \citet{van12} has shown that the reddening of the host galaxy is $E(B-V)=0.08$, but \citet{koc12} suggest that it is even less. 

\citet{gal08} suggest that all SNe Type~II-P should be very similar, which is seen here with the SN 2012aw spectrum falling largely in between the two models. According to \citet{des08}, the main difference between SN 2006bp and SN 2005cs is that SN 2005cs may be intrinsically faint with low kinetic energy in the ejecta.  However, it is not surprising that there some are discrepancies between the models and the observations of SN 2012aw.  The best UV spectral models are based on early time spectra and late time photometry, including these. Now that there are late time UV spectral observations of a normal Type II-P (versus the peculiar SN 1987A), improvements to the spectral models can be made.  

\section{Conclusions}
SN 2012aw shows the same optical light curve behavior as other Type~II-P SNe, namely an optical plateau due to the reprocessing of hydrogen recombination emission.   This SN was also UV bright early on and faded rapidly in the same manner as other observed Type~II-P SNe. However, at late epochs, after Day 27, the light curve slope flattened to a much slower decline.  At these later times, the UV lightcurve uniquely gives information on temperature changes. Thus, the UV flattening indicates that the temperature is remaining constant, which requires a source of heating to maintain the photosphereic temperature.  Some possible sources of late-time heating are continued recombination, radioactive decay, or a combination of these. 

Past Day 27, as the UV slope flattens, the ${\it uvm2}-{\it v}$ color flattens and the temperature flattens.  The bolometric temperature as calculated from the UVOT filters {\it uvw1} and {\it uvw2} and ground-based UBVRIJHK (Dall'Ora et al., in prep) becomes constant at $\sim$ 4500~K. A blackbody temperature of 4500~K would have a color of ${\it uvm2}-{\it v} = 4.1$ mag including the \citet{van12} reddening, but we observe ${\it uvm2}-{\it v} \sim 6$ mag. The difference in color is likely due to line blanketing.  We note, however, that there have been some indications of a higher reddening value from Dall'Ora et al. (in prep) if there is significant dust present after the explosion.  They suggest an $E(B-V)=0.27$, which would produce a 4500~K blackbody color of ${\it uvm2}-{\it v} = 5.3$.  If this is so, then line blanketing may be less significant and the SN was intrinsically brighter.

The  UV plateau in SN 2012aw is not seen in the other UV observed SNe because most were either too extincted (SN 2006bc: $E(B-V)=0.19$ -- line of sight \citealt{sch11}; $E(B-V)=0.33$ -- host \citealt{ots12}) or too far away (SN 2007aa: 20.5 Mpc, \citealt{sma09}; SN 2006at: 64 Mpc, \citealt{blo06}).  SN 2005cs was nearby (in M51) and not very extincted, but \citet{des08} suggest that it was sub-luminous.  SN 2006bp is the best comparison and has indications that the lightcurve was beginning to flatten, but it too was fairly distant and extincted (17.5 Mpc, $E(B-V)=0.37$ host; \citealt{des08}) and was not detected in the UV approximately a month after explosion.  Thus,  we have lengthy observations of a normal Type II-P SNe that is nearby allowing us to observe the UV plateau for the first time.
 
Because of SN 2012aw seeming similarity to SN 2006bp and at least the initial data of the other SNe, this UV plateau phase could be ubiquitous.\footnote[3]{We note that galactic template subtractions are sometimes done with post-explosion data.  We caution that there may be SN UV flux even at late times in low-extinction SNe.}  However, as SN 2012aw is the first we can only confirm this with additional observations of nearby Type II-P SNe.

\acknowledgments
The authors would like to thank Luc Dessart for providing the digital models for SN 2005cs and SN 2006bp. This research has made use of the NASA/IPAC Extragalactic Database (NED) which is operated by the Jet Propulsion Laboratory, California Institute of Technology, under contract with the National Aeronautics and Space Administration. The authors would like to thank the anonymous referee for their insightful comments. 
  \newpage
  

\end{document}